\documentclass[prl,aps,twocolumn,showpacs,preprintnumbers]{revtex4}

\usepackage{psfrag,graphicx}
\usepackage{dcolumn}
\usepackage{amsmath,amssymb}
\usepackage{bm}
\usepackage{amsfonts,amssymb,amsmath}        
\usepackage{epstopdf}

\newcommand{\be}{\begin{equation}}
\newcommand{\ee}{\end{equation}}
\newcommand{\bq}{\begin{eqnarray}}
\newcommand{\eq}{\end{eqnarray}}

\newcommand{\no}{\nonumber\\}

\begin{document}

\title{Cold atom simulation of interacting relativistic quantum field theories}

\author{J. Ignacio Cirac$^1$, Paolo Maraner$^2$ and Jiannis K. Pachos$^{3}
\footnote{Email: j.k.pachos@leeds.ac.uk}
$}
\affiliation{$^1$Max-Planck-Institut f\"ur Quantenoptik, Hans-Kopfermann-Str. 1, D-85748 Garching, Germany}
\affiliation{$^2$School of Economics and Management, Free University of Bozen-Bolzano, via Sernesi 1, 39100 Bolzano, Italy}
\affiliation{$^3$School of Physics and Astronomy, University of Leeds, Woodhouse Lane, Leeds LS2 9JT, UK}
\date{\today}
\begin{abstract}

We demonstrate that Dirac fermions self-interacting or coupled to dynamic scalar
fields can emerge in the low energy sector of designed bosonic and fermionic cold atom systems. We illustrate this with two examples defined in two spacetime dimensions. The first one is the self-interacting Thirring model. The second one is a
model of Dirac fermions coupled to a dynamic scalar field that gives rise to the
Gross-Neveu model. The proposed cold atom experiments can be used to probe spectral or correlation properties of interacting quantum field theories thereby presenting an alternative to lattice gauge theory simulations.

\end{abstract}

\pacs{11.15.-q, 11.15.Ha}

\maketitle

{\bf Introduction:-} To date there exist several important open problems in
quantum field theories (QFTs) ranging from  the
convergence of the scattering matrix to the infrared behaviour of quantum
chromodynamics (QCD). Such problems can be probed analytically only by
non-perturbative methods that seem to be intractable in three and four
dimensions. As an alternative, numerical techniques or quantum simulations can play a
central role in obtaining insight into the Standard Model.

Recently, much interest is focused on simulating QFTs with cold
atoms~\cite{Haller,Buchler, Kormos,Semiao,Bermudez}. In the relativistic domain, these
studies are usually limited to generating Dirac
fermions~\cite{Gerritsma,Vaishnav} and background
fields~\cite{Ruostekoski,Lewenstein,Maraner}. In this letter, we propose the
realisation of Dirac fermions interacting with {\em dynamic} fields by
employing cold atoms in optical lattices. This constitutes the first step
towards the quantum simulation of a general model with coupled relativistic fermionic
and bosonic fields. Cold atoms provide a controlled medium with well
understood interactions. A wide range of quantum optics and atomic
physics technology allows for the preparation, manipulation and detection of a variety of
interesting many body phenomena. Employing cold atoms gives us the possibility to consider two,
three and four spacetime dimensions, to tune the couplings of the
interactions and to explore the behavior of multi-component fields. Compared
to numerical simulations of lattice gauge theories on computers, a
physical simulation on a quantum system naturally overcomes the sign
problem~\cite{Baeurle}.

Here we show how to simulate a two-dimensional self-interacting model of
Dirac fermions, known as the Thirring model~\cite{Thirring,Korepin} and
two-dimensional Dirac fermions coupled to a scalar field that is equivalent
to the Gross-Neveu model. The Hamiltonians of these systems are supported on
one spatial dimension. The necessary building blocks are the Dirac
Hamiltonian, which describes relativistic fermions, and the interaction of
fermions with themselves or with a dynamic scalar field. This goes beyond previous proposals concerned with dynamical fermions coupled to classical fields. We show how the required
components naturally emerge in the low energy sector of specifically
designed lattice Hamiltonians. The Dirac operator describes the continuum
limit of certain fermionic lattices~\cite{Susskind,Jackiw,Ruostekoski}, as
in graphene~\cite{Neto}, that remarkably give rise to both the spin and the
linear dispersion relation. Compared to previous
approaches~\cite{Ruostekoski} here we realise the Dirac operator only by the
free tunnelling of single species of atoms. Slow spatial variations of the
lattice couplings result, in the Dirac picture, to an interaction with a
scalar field background~\cite{Maraner}. When these distortions are caused by
coupling the lattice fermions to bosons, the resulting fields become
dynamic. Self-energy terms can be implemented giving rise to a variety of
interesting QFTs. The presented models are exactly solvable and serve for
demonstrating the ability to simulate important properties of the Standard
Model such as dynamical symmetry breaking and mass generation with cold
atoms.


{\bf Thirring model:-} Our starting point is the two-dimensional Thirring
model~\cite{Thirring,Korepin}. It describes interacting fermions with the
Hamiltonian
\be
{H_\text{T}\over \hbar}=\int dx \Big(v_\text{s}\bar \Psi \gamma_1p \Psi
+{m_0v_\text{s}^2 \over \hbar} \bar\Psi\Psi +{g
\over 2}\bar\Psi\gamma^\mu\Psi\bar\Psi\gamma_\mu\Psi\Big).
\label{Thirring}
\ee
Here the $\gamma$'s are two-dimensional matrices satisfying $\{\gamma_\mu,\gamma_\nu\}=2\eta_{\mu\nu}$ for $\mu,\nu=0,1$ with $\eta_{\mu\nu} =\text{diag}(1,-1)$, $\gamma_5=\gamma_0\gamma_1$ and $\Psi$ is a two-dimensional spinor with $\bar \Psi =\Psi^\dagger \gamma_0$. The mass of the fermions is $m_0$, $g$ is
their dimensionless self-interaction coupling constant and $v_\text{s}$ is
the sound velocity taken in high energy to be the speed of light. This model
has exciting physics with the massless case, $m_0=0$, being equivalent to free
bosons and the massive case being equivalent to the Sine-Gordon
model~\cite{Coleman}.

\begin{figure}[t]
\begin{center} {\includegraphics[width=8cm]{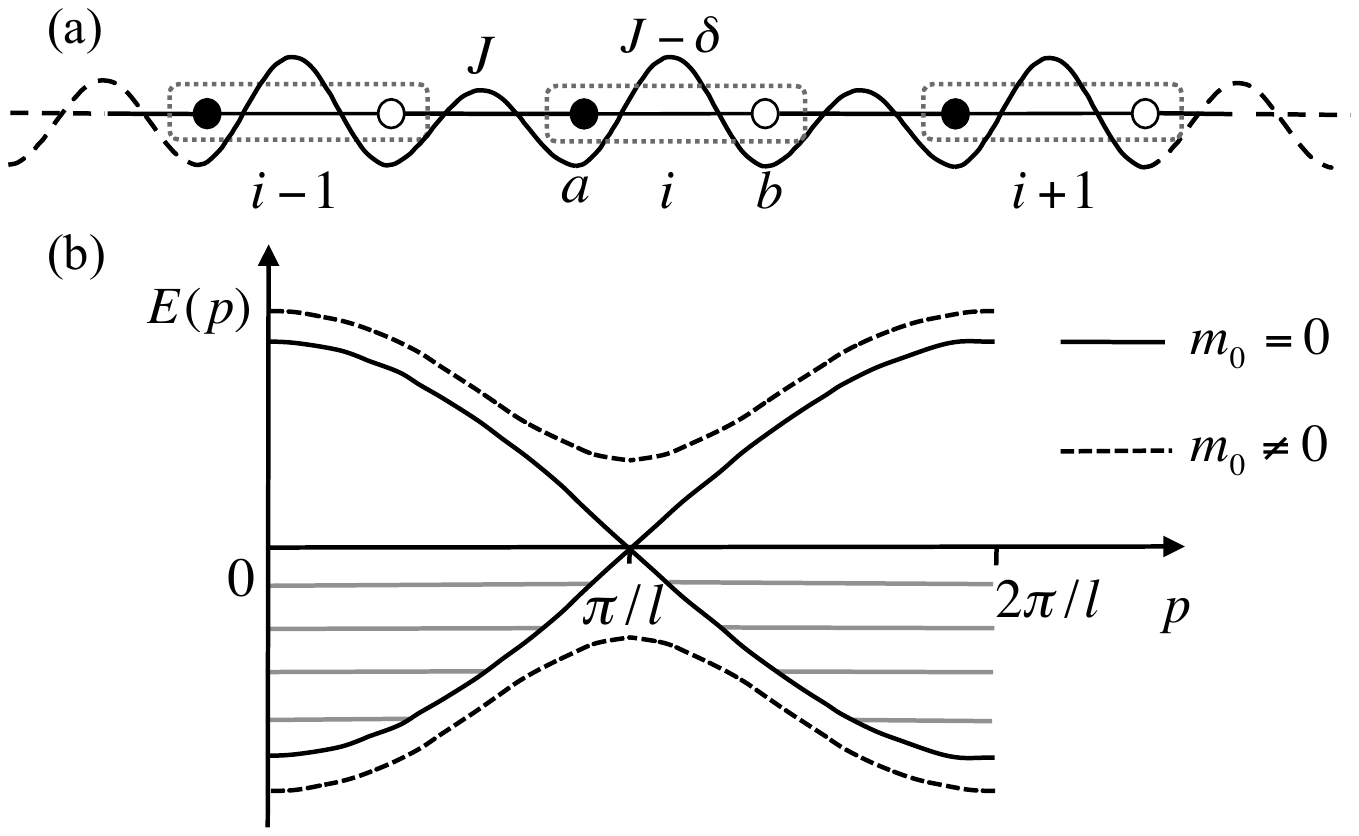}}
\caption{\label{figure1} (a) The one-dimensional superlattice with tunnelling
fermionic atoms that simulates Dirac fermions. Each unit
cell includes two fermion sites, $a$ and $b$. An alternating distortion of the
tunnelling couplings $J$ and $J-\delta $ gives
rise to the mass term, $\delta $. (b) The energy dispersion relation as a
function of momentum, $p$. At half filling and for $\delta =0$ the low
energy behaviour is linear with respect to $p$, allowing the Dirac operator description.}
\vspace{-0.5cm}
\end{center}
\end{figure}

To simulate the Thirring model in an atomic system we consider a linear
bicolourable fermionic lattice with spacing $l$ (see Fig.~\ref{figure1}(a)) which is
subject to the Hubbard Hamiltonian
\bq
{H \over \hbar}=2\pi\sum_i &&\!\!\!\!\!\Big[-J(a_i^\dagger b_i
+ b_i^\dagger a_{i+1})+\delta a_i^\dagger b_i +\text{H.c.}
\no\no&&
+Ua_i^\dagger a_ib^\dagger_ib_i\Big].
\label{ThirringLattice}
\eq
Here $J$ is the tunnelling coupling between neighbouring sites of the
lattice of the same fermionic atoms $a$ and $b$ with
$\{a_i,a^\dagger_j\}=\{b_i,b^\dagger_j\}=\delta_{ij}$ and all the other
anticommutators vanishing. The $a$, $b$ index is a spatial distinction
within the unit cell that allows the encoding of the spin degree of freedom.
The tunnelling distortion $\delta$ occurs on alternating links,
as shown in Fig.~\ref{figure1}(a), while $U$ is the interactions coupling
between fermions in the same cell.

If we diagonalise the $J$-term of the Hamiltonian in the $a$, $b$ basis
we find the dispersion relation $E_\pm (p) = \pm 2|\cos {pl \over 2}|$
plotted in Fig.~\ref{figure1}(b). It can be easily seen that there is a
single Fermi point, $P=\pi/l$, for which $E_\pm(P)=0$. If the lattice is
half filled with fermions, which occupies the valence band completely, the
behaviour of the small energy fluctuations is governed by the Hamiltonian
expanded around $P$. Setting $p=P+k$ for $|k|\ll1/l$ we obtain a dispersion
that is linear in momentum, $k$. Hence, the fermionic tunnelling term around
the Fermi point can be efficiently described by the relativistic Dirac
Hamiltonian $J l\int dk\psi^\dagger_k\sigma_2 k\psi_k$, where $\psi=(a,b)^T$
corresponds to the one-dimensional version of the Kogut-Susskind
fermions~\cite{Susskind,Jackiw}. To assign the appropriate dimensions to the
Dirac fermions we set $\psi =\sqrt{2l}\Psi$. In this way we obtain the
massless free Dirac fermion term of (\ref{Thirring}) written in the
momentum representation with $\gamma_0=\sigma_1$, $\gamma_1=i\sigma_3$ and
$v_\text{s}=2\pi \cdot 2lJ$. It can be verified that the continuum
limit of (\ref{ThirringLattice}) gives also rise to the rest of the terms in
(\ref{Thirring}) with ${m_0v_\text{s}^2/
\hbar} = 2\pi \cdot 2\delta$ and ${g} =2\pi \cdot 2Ul$.
The continuum limit corresponds to small lattice spacing. This is equivalent to restricting to the low energy sector of the system where states have a large wavelength support. These are exactly the states we are interested in for probing the infrared behaviour of QFT, such as the ground state and its gapped or gapless nature.

The Hamiltonian (\ref{ThirringLattice}) can be realised with cold atoms as
follows. The one-dimensional fermionic tunnelling term of
(\ref{ThirringLattice}) appears when a fermionic gas is placed in an optical
lattice with very tight confinement in the other two directions. The
$\delta$-term corresponds to a uniformly decreased tunnelling coupling
between sites of the same cell. It can be generated, e.g. by employing superlattices as seen in Fig.~\ref{figure1}(a). This alternatively signifies that inhomogeneity in the tunnelling coupling due to experimental imperfections will generate a mass term as observed in~\cite{Haller}. The final
$U$-term results from the interaction between the atoms $a$ and $b$ present
in the same cell [See supplementary material at [URL will be inserted by AIP]].

\begin{figure}[t]
\begin{center} {\includegraphics[scale=.3]{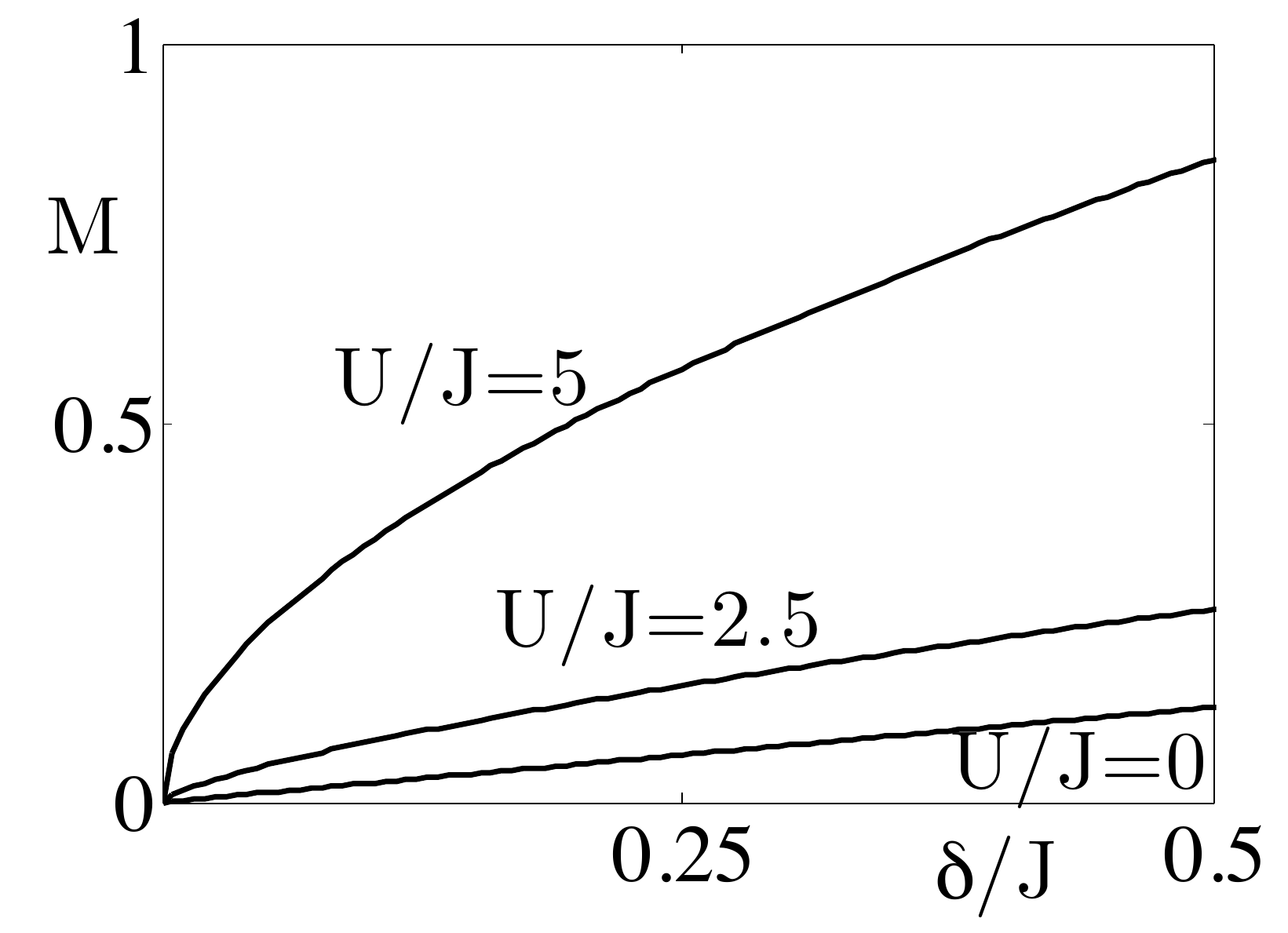} }

\caption{\label{figure2} The regularised mass $M$, in units of $\hbar/(v_\text{s} l)$,
as a function of
the tunnelling disorder $\delta $ for various interaction strengths $U$.
When no interactions are present the mass increases, as expected, linearly as a function
of $\delta $. The presence of interactions dramatically changes
this behaviour even for moderate ratios of $U/J$. }
\vspace{-0.5cm}
\end{center}
\end{figure}

The Thirring model is the simplest relativistic interacting QFT that one
could implement in the laboratory with present technology. Its realisation can
demonstrate the renormalisation of mass due to interactions. Indeed,
$m_0\neq 0$ is the fermionic mass of the classical theory. When the
interactions $g$ are introduced then a regularisation condition needs to be
adopted, $|p|<\Lambda$, where $\Lambda$ is a momentum cutoff that excludes
modes with unphysically high energy. Our system is naturally regularised due
to the underlining lattice structure, where the momentum cutoff is related
to the optical lattice spacing by, $\Lambda = {\pi / l}$. A quantum field theory is
called renormalisable when this cutoff can be absorbed in the initial, bare
parameters of the model, such as the mass. The massive Thirring model is
such a renormalisable theory where the interactions $g$ give rise to the
following regularised mass
\be
{M v_\text{s} \over \hbar} = {\delta \over l}  {\tan {\pi U\over \pi J +U}\over U}
\exp ({U \over \pi J+U} \sinh^{-1} {\pi J\over \delta}).
\ee
The corresponding coupling regime $0\leq U/J<\pi$ is well within the realm
of current experiments~~\cite{Jordens}. This effective mass is exactly the
energy gap above the ground state obtained for zero momentum excitations,
while non-zero momenta give a continuum spectrum above the gap. Fixing the
lattice spacing, $l$, according to the wavelength of the optical lattice,
one can plot the effective mass as a function of the atomic parameters
$\delta$ and $U$, as seen in Figure~\ref{figure2}. Current experiments
routinely probe such excitation gaps in cold atom systems~\cite{Haller}.

The atomic simulation relies on direct exchange of atoms, while the value of $M$ increases when the couplings $U$
or $\delta $ increase, in a non-perturbative way, or when the lattice spacing $l$ decreases. The size of
the gap can, hence, be efficiently controlled with a variety of experimental means which alleviates the low temperature requirements \cite{Bloch}. This facilitates
the experimental measurement of the gap by a spectral analysis of the atomic
system~\cite{Haller}. The renormalisation step would then necessitate to
take the `bare' mass $m_0$ decrease as a function of $\Lambda$ so that a
finite constant value of the `physical' mass is obtained. It is intriguing
that this renormalisation procedure can be established experimentally by
studying the spectral behaviour of the fermionic lattice system. Observing
such a strong renormalisation of parameters (see Fig.~\ref{figure2}) in a
system of cold atomic gases provides a unique fingerprint of strong
correlations.


{\bf Fermion-scalar interaction and the Gross-Neveu model:-} Next we
consider a two-dimensional model where an $N$-colour massless Dirac fermion
$\Psi_n$ $n=1,...,N$ interacts with a massive quantised scalar field $\Phi$
according to the Hamiltonian
\be
{H_\Phi \over \hbar} = \int dx \Big(v_\text{s}\bar\Psi_n  \gamma_1 p\Psi_n+
g m \Phi\bar\Psi_n\Psi_n+{m^2 \over 2 }\Phi^2 \Big).
\label{Hscalar}
\ee
Here we assume summation over the colour index $n$, $g$ is the coupling
strength between the bosonic and fermionic fields and $m$ is a mass scale
that can be absorbed in $\Phi$. It can be shown that this Hamiltonian yields
exactly equivalent fermionic behavior as the Gross-Neveu model~\cite{Gross}
given by
\be
{H_\text{GN}\over \hbar}=\int dx \Big[v_\text{s}\bar \Psi_n \gamma_1p \Psi_n
+{g^2\over 2}(\bar\Psi_n\Psi_n)^2\Big].
\label{Gross-Neveu}
\ee
The Hamiltonian $H_\text{GN}$ describes massless fermions subject to
attractive interactions. This attraction causes the fermions to create
bosonic pairs. Eventually, these composite bosons condense breaking
spontaneously the $Z_2$ symmetry, $\Psi\rightarrow \gamma_5\Psi$, of
Hamiltonian (\ref{Gross-Neveu}), thereby causing the fermions to dynamically
acquire mass. This fascinating property is very similar to the behaviour of
the BCS theory of superconductivity or of four-dimensional QCD. Note that
(\ref{Hscalar}) does not contain a kinetic term for the scalar field. This
corresponds to a Yukawa theory with interactions mediated by infinitely
massive fields, making their propagation point-like and resulting in the
Gross-Neveu effective four-fermion interaction.

We will now consider a cold atom system that gives rise to
$H_\Phi$~\cite{Comment1}, which would make it possible to observe the
dynamical mass generation experimentally. The kinetic term of the Dirac
fermions, $\Psi_n$, can be produced by the same fermionic tunnelling term as
in (\ref{ThirringLattice}). In general, a variety of interaction terms can
be generated between bosonic and fermionic atoms. To conform with
(\ref{Hscalar}), we specifically want the bosonic modes, $\Phi$, to couple
linearly to the fermionic ones, $\Psi_n$, as dictated by the minimal
coupling prescription. Such an interaction can result similarly to the
$m_0$-term of the Thirring model. A position dependent tunnelling distortion
gives rise to a classical scalar field configuration. Formally, the
quantisation of the scalar field is obtained by writing it as $\Phi =
(d^\dagger +d)/\sqrt{2}$, where $d$ is a bosonic mode. Substituting it into
$H_\Phi$ gives the quantised Dirac fermion-scalar model. In the cold atom
setting, this can be achieved by employing a bosonic condensate interacting
with the lattice fermions as we shall see below.

\begin{figure}[t]
\begin{center} {\includegraphics[width=8cm]{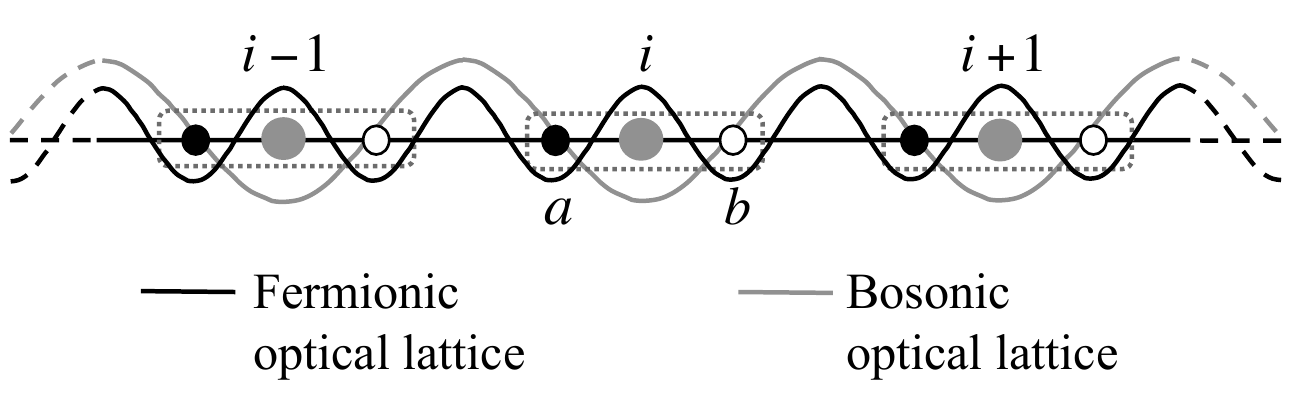} }
\caption{\label{figure3} The one-dimensional optical lattice with tunnelling fermions and
bosons that simulates the Dirac fermion-scalar field model. The bosonic
sites are placed in-between the fermionic ones with double spacing. In this
way the bosonic population on site $i$ controls the fermionic tunnelling
within the same cell. }
\vspace{-0.5cm}
\end{center}
\end{figure}

Consider a one-dimensional fermionic
lattice superposed with a one-dimensional bosonic lattice, as seen in
Fig.~\ref{figure3}. We assume that the dynamics of the atoms is described by
\bq
&&{H \over \hbar} =2\pi \sum_i
\Big\{\big[-(J'a_{n,i}^\dagger b_{n,i} +J b_{n,i}^\dagger
a_{n,i+1}) -\no\no &&
\,\,\,\,\,\delta\alpha_i^\dagger
\alpha_i a_{n,i}^\dagger b_{n,i} +\text{H.c.}\big]+U
{\alpha_i^\dagger}^2 {\alpha_i}^2 + \mu
\alpha_i^\dagger \alpha_i\Big\}.
\label{atoms1}
\eq
Here the $a_{n}$'s and $b_{n}$'s are $N$ different species of fermionic
atoms (summation over $n$ is assumed) and $\alpha_i = D+d_i$ is an atomic
condensate with particle density $D$ and bosonic operator $d$. When the
couplings of (\ref{atoms1}) and the condensate density are appropriately
tuned, then the low energy behaviour of the Hamiltonian $H$ reproduces
$H_\Phi$. Indeed, for $J'=J - \delta D^2 $ and $\mu = -2D^2 U$ one reproduces
the desired low energy behavior with $g^2=2\pi
\cdot 2 \delta^2l/U$ and $v_\text{s}=2\pi\cdot 2J l$. To suppress
spurious terms we take the weak fluctuation limit $\langle d^\dagger
d\rangle \ll D^2$.

The terms in (\ref{atoms1}) can be realised in the lattice configuration of
Fig.~\ref{figure3} by employing one-dimensional optical lattices filled with
the appropriate species of atoms. All the required interactions naturally
appear in cold atom settings and can be tuned, e.g. by Feshbach resonances.
The $J$- and $J'$-terms of (\ref{atoms1}) result from the tunnelling of the
fermionic atoms along their lattice and contribute to the free Dirac fermion
propagation. The $\delta$-term results from the interaction between bosonic
and fermionic atoms that generate a fermionic tunnelling controlled by the
bosonic populations. It gives rise to the interaction between the scalar and
the fermion fields. In (\ref{atoms1}) there is no tunnelling term for the
bosonic atoms. Hence, the required system comprises of independent BECs at
each site of a bosonic lattice~\cite{Schweikhard,Hadzibabic}. The $U$-term
describes the interaction between bosonic atoms on the same site and the
$\mu$-term is given by their chemical potential. Importantly, small errors
in the values of all of these couplings result in modifications of the
parameters of the effective Hamiltonian (\ref{Hscalar}) or the generation of
similar terms~\cite{Maraner}. The persistence of the effective Dirac
description in the low energy limit is a characteristic also present in the
Thirring model simulation.

It has been shown~\cite{Gross} that the initially massless fermions of the
Gross-Neveu model dynamically acquire an effective mass. In terms of the
atomic parameters, the emerging mass gap is given by
\be
{Mv_\text{s} \over \hbar} = {\pi \over l} \exp(-{\pi \over N} {J U \over \delta^2}).
\label{mass2}
\ee
This result is exact in the large $N$ limit but is expected to be a good
approximation even for moderate, experimentally feasible, values of $N$ of the order of $2$ or $3$. While the mass, $M$, goes to infinity when
$l$ goes to zero, the renormalisation prescription absorbs the infinity in
the `bare' coupling $g$. In the case of the optical lattice realisation we
do not need to renormalise the coupling $g$ as we are working with a  fixed
lattice spacing $l$. Such a simulation can detect the dynamical generation
of mass and verify the predicted behaviour (\ref{mass2}) as a function of
the atomic couplings.


{\bf Conclusions:-} The above method can be straightforwardly generalised to
other more complex QFTs. For example, to simulate quantum
electrodynamics in two spacetime dimensions (QED$_2$) we need a similar atomic setting
to the fermion-scalar model described above. Apart from the addition of a
kinetic term for the bosonic field we need to impose the Gauss constraint in
the atomic level. An atomic model that can give rise to QED$_2$ as well as its
generalisation to four dimensions will be presented elsewhere.

Going beyond QED one could realise the SU$(N)$ Yang-Mills gauge field
coupled to an $N$-colour Dirac fermion (see (\ref{Hscalar})) that gives rise
to QCD. The challenge faced at this point is to realise the higher order
bosonic self-interaction terms that are necessary to simulate the non-linear
behaviour of the Yang-Mills theory. It is a fascinating perspective to probe
the generation of a mass gap in such a QCD simulation. In parallel, a wealth
of possibilities opens for the simulation of supersymmetric QFTs or
combining QFT and relativity in an atomic system. It is highly plausible
that the experimental realisation of the simple models presented here
provide a unique probing tool into the open questions of interacting
relativistic QFTs.

{\bf Acknowledgements:-} J.K.P. would like to thank N. Cooper, J. Dunningham and S. Ruijsenaars for inspiring discussions. This work was supported by the Royal Society. J.I.C. acknowledges funding from the EU (project AQUTE) and the DFG (Forchergruppe 635).


\begin{thebibliography}{99}
\bibitem{Comment1}
Hamiltonian (\ref{Hscalar}) is significantly simpler to simulate than the
actual Gross-Neveu one (\ref{Gross-Neveu}) as the interaction term in the latter
has a summation over the fermionic components squared that demands the fine
tunning of many atomic interactions.
\end{thebibliography}
\end{document}